\documentclass{ws-ijmpd}
\usepackage{graphicx,amssymb,amsmath,amsfonts}
\usepackage{epsfig}
\begin{document}

\markboth{Parthapratim Pradhan}{(Thermodynamic Products in Extended Phase Space)}

\title{Thermodynamic Products in Extended Phase Space }
\author{Parthapratim Pradhan\footnote{pppradhan77@gmail.com.}}

\address{Department of Physics, Vivekananda Satavarshiki Mahavidyalaya (Affiliated to Vidyasagar University),
West Midnapur, West Bengal~721513, India}

\maketitle

\begin{history}
\received{Day Month Year}
\revised{Day Month Year}
\comby{Managing Editor}
\end{history}

\begin{abstract}
We have examined the thermodynamic properties for a variety of spherically symmetric charged-AdS 
black hole (BH) solutions, including the charged AdS BH surrounded by quintessence dark energy 
and charged AdS BH in $f(R)$ gravity in \emph{extended phase-space}. This framework involves treating 
the cosmological constant as thermodynamic variable (for example: thermodynamic pressure and 
thermodynamic volume). Then they should  behave as an analog of Van der Waal (VdW) like 
systems. In the extended phase space we have calculated the \emph{entropy product} and 
\emph{thermodynamic volume product} of all horizons. The mass (or enthalpy) independent 
nature of the said product signals they are \emph{universal} quantities. 
The divergence of the specific heat 
indicates that the second order phase transition occurs under certain condition. In the 
appendix-A, we have studied the thermodynamic volume  products for axisymmetric spacetime 
and it is shown to be \emph{not universal} in nature. Finally, in appendix-B, we have studied 
the $P-V$ criticality of Cauchy horizon for charged-AdS BH and found to be an universal relation 
of critical values between two horizons as $P_{c}^{-} = P_{c}^{+}$, $v_{c}^{-}=v_{c}^{+}$, 
$T_{c}^{-} = -T_{c}^{+}$, $\rho_{c}^{-} = -\rho_{c}^{+}$. The symbols are defined 
in the main work.

\end{abstract}


\section{Introduction}
An interesting topic in recent years both in the general relativity community and in the String theory community is 
that the BH area (or entropy) product formula of all horizons independent of 
the ADM (Arnowitt-Deser-Misner) mass of the background space-time 
\cite{ah09,cgp11,castro12,chen12,sd12,mv13,castro13,ppepjc,ppplb,ppsen}. 
For example, the area product formula for a Kerr BH \cite{ah09} depends only on the angular momentum parameter:
\begin{eqnarray}
{\cal A}_{2} {\cal A}_{1} &=& 64\pi^2J^2 ~.\label{prKN}
\end{eqnarray}
where ${\cal A}_{2}$ and ${\cal A}_{1}$ are  area of the inner and outer horizons.

Whenever, we have taken the perturbed space-time with a spinning BH in some non-trivial 
environment e.g. a BH surrounded by a ring of matter or a multiple BH space-time the same 
formula holds. Hence, the area product formula of outer horizon or event horizon (${\cal H}^{+}$) 
and inner horizon or Cauchy horizons (${\cal H}^{-}$) for Kerr BH is of an \emph{universal} 
quantity: it holds independently of the environment of the BH.

On the other hand, if we incorporate the BPS (Bogomol'ni-Prasad-Sommerfeld) states, the area product 
formula of ${\cal H}^{\mp}$ should read \cite{cgp11}
\begin{eqnarray}
{\cal A}_{2}{\cal A}_{1} 
&=& 64 \pi^2 {\ell _{pl}}^4 N , \,\, N\in {\mathbb{N}}, N_{1}\in {\mathbb{N}}, N_{2} \in {\mathbb{N}}
~.\label{ppl}
\end{eqnarray}
where $\ell _{pl}$ is the Planck length. This indicates the area product should be quantized. 

Alternatively, the area products independent of mass  implies that there should be an important role 
of the Cauchy horizon in the BH thermodynamics as well as in BH physics. Now the relevant question is 
that the mass independent product formula is generic? It has been shown explicitly by Visser \cite{mv13}that 
by incorporating the cosmological constant, the area product of all physical horizons is not mass 
independent. But typically, some complicated function of inner and outer horizon area is indeed mass 
independent. 

Previous studies have not made use of the extended phase space formalism. 
Thus in this work, we wish to examine the thermodynamic product formula in \emph{extended
phase space}. Where the ADM mass of an AdS BH could be treated as the enthalpy of 
the space-time and the cosmological constant should be treated as the thermodynamic 
pressure \cite{kastor09}. Therefore there must exists a conjugate quantity 
which is a thermodynamic volume associated with the BH space-time.

Besides area (or entropy) products, it needs to be evaluated whether other thermodynamic 
products \cite{ppepjc,ppplb,ppmpla,ppahep,ppjetpl,ppjetp} like BH temperature products, specific heat products, Komar energy 
products etc. are provide any universal characterization or not, and here we first introduced 
the \emph{thermodynamic volume products} when one must considered the extended phase space thermodynamics. Does it 
independent of the ADM mass parameter? We will investigate this issues 
in the present work. So, when the cosmological constant treated as a thermodynamic pressure 
and its conjugate variable as a thermodynamic volume what happens the \emph{Smarr mass formula}, 
\emph{Smarr-Gibbs-Duhem} relation  and BH \emph{equations of state} in the extended phase space.   Additionally, we 
find the mass independent volume products relation in parallel with the entropy product relations.

BH thermodynamic properties have been investigated for many decades and still it is going on. In the present study, the 
main motivation comes from the seminal work of Hawking and Page \cite{haw83} where the thermodynamic properties of BHs in 
Schwarzschild-AdS space has been explicitly studied. The author discussed the phase transition (between small and large BHs for 
Schwarzschild-AdS BH) which is called famous Hawking-Page phase transition. The special interest is due to the 
application of AdS space-time in gauge/gravity duality via dual conformal field theory (CFT) through AdS/CFT 
correspondence \cite{witten}. Several exotic phenomena occurs in the AdS space-time. First example of course 
be Hawking-Page phase transition in Schwarzschild-AdS spacetime. The second one is that in charged AdS spacetime, the 
gravitational analogue of the liquid/gas phase transition has been observed in the phase diagram which was explicitly 
investigated  by several authors \cite{chamblin99,chamblin99a,gubser,david12} and the fact that for charged AdS BH the 
notion of thermodynamic equilibrium is a straightforward concept. The third one is that Kerr-AdS spacetime admits reentrant 
phase transition and showing a tri-critical point in their phase diagram \cite{david13}.

The current interest is involved due to the variation of negative cosmological constant and also it is proportional to the 
thermodynamic pressure \cite{dolan10,dolan11,david12}.  The thermodynamic products especially area (or entropy) 
products in charged AdS BHs  were calculated in \cite{mv13} but the author has not been considered there the 
extended phase space. Here we shall compute the volume products by considering the thermodynamic pressure ($P$) 
is equal to the negative cosmological constant ($\Lambda$) divided by $8\pi$ (where $G=c=k=\hslash$=1) i.e.
\begin{eqnarray} 
P &=& -\frac{\Lambda}{8\pi}=\frac{3}{8\pi \ell^2}  ~.\label{pr}
\end{eqnarray}
and the corresponding thermodynamic volume can be defined as 
\begin{eqnarray}
V &=& \left(\frac{\partial M}{\partial P}\right)_{S,Q,J}  ~.\label{vm}
\end{eqnarray} 
This volume for charged-AdS BH should read as 
\begin{eqnarray}
V_{i} &=& \frac{4}{3}\pi r_{i}^3  ~.\label{vm1}
\end{eqnarray} 
where $r_{i}$ is the corresponding horizon radius and $i=1,2,3,4$.

It has been shown that the \emph{Reverse Isoperimetric Inequality} is satisfied for 
event horizon \cite{cvetic11}. Here we conjecture that this inequality is valid for 
all the horizons i.e.
\begin{eqnarray}
 {\cal R}_{i} &=& \left(\frac{3V_{i}}{4\pi} \right)^{\frac{1}{3}} 
 \left(\frac{4\pi}{{\cal A}_{i}} \right)^{\frac{1}{2}} \geq 1 ~.\label{rip}
\end{eqnarray}
It should be noted that a class of BHs with non-compact event horizons do not satisfy 
this inequality \cite{mannprl,mannjhep}.

The structure of the paper is as follows. In Sec.(\ref{tnt}), we have described the thermodynamic properties 
of RN-AdS BH. The Sec.(\ref{quint}) describes the thermodynamic properties of the RN-AdS BH surrounded by 
quintessence. In Sec. (\ref{fr}), we have given the thermodynamic properties of the
 $f(R)$ gravity. Finally, we conclude in Sec.(\ref{dis}). In appendix-A, we have examined  the thermodynamic 
volume  products for axisymmetric space-time and in appendix-B, we have investigated the $P-V$ criticality 
of inner horizon for RN-AdS BH.

\section{\label{tnt} Thermodynamic properties of Charged AdS BH:}
Let us begin with the charged-AdS space-time metric can be written as in terms of 
Schwarzschild like coordinates\cite{chamblin99,chamblin99a}:
\begin{eqnarray}
ds^2 &= & -{\cal U}(r) dt^2 + \frac{dr^2}{{\cal U}(r)} +r^2 d\Omega_{2}^2 .~\label{metric}
\end{eqnarray}
where,
\begin{eqnarray}
{\cal U}(r) &=&  1-\frac{2M}{r}+\frac{Q^2}{r^2}+\frac{r^2}{\ell^2},\label{h1}
\end{eqnarray}
and $d\Omega_{2}^2$ is the metric on the unit sphere in two dimensions.

The electromagnetic potential one form for the space-time (\ref{metric}) is
\begin{eqnarray}
A=A_{\mu}dx^{\mu}=-\frac{Q}{r}dt.~\label{h2}
\end{eqnarray}

The BH horizons determined by the condition ${\cal U}(r)=0$ i.e. 
\begin{eqnarray}
\frac{r^4}{\ell^2}+r^2-2Mr+Q^2 &=& 0 ~.\label{rna1}
\end{eqnarray}
In terms of thermodynamic pressure, this could be rewritten as 
\begin{eqnarray}
\frac{8\pi P}{3}r^4+r^2-2Mr+Q^2 &=& 0 ~.\label{rna2}
\end{eqnarray}
To finding the roots we apply the Vieta's rule, we get 
\begin{eqnarray}
\sum_{i=1}^{4} r_{i} &=& 0 ~.\label{eq1}\\
\sum_{1\leq i<j\leq 4} r_{i}r_{j} &=& \frac{3}{8\pi P} ~.\label{eq2}\\
\sum_{1\leq i<j<k\leq 4} r_{i}r_{j} r_{k} &=&  \frac{3M}{4\pi P} ~.\label{eq3}\\
\prod_{i=1}^{4} r_{i} &=&  \frac{3Q^2}{8\pi P} ~.\label{eq4}
\end{eqnarray}
The entropy of the BH can be defined as 
\begin{eqnarray}
{\cal S}_{i}  &=& \frac{{\cal A}_{i}}{4} ~.\label{eq5}
\end{eqnarray}
where the area of the BH is 
\begin{eqnarray}
{\cal A}_{i}  &=& 4\pi r_{i}^2 ~.\label{eq6}
\end{eqnarray}
and now the BH temperature reads as 
\begin{eqnarray}
T_{i} &=& \frac{{\cal U}'(r)}{4\pi}= 
\frac{1}{4\pi r_{i}} \left(1+8\pi P r_{i}^2-\frac{Q^2}{r_{i}^2} \right)~. \label{eq7}
\end{eqnarray} 
The electric potential could be defined as 
\begin{eqnarray}
\Phi_{i}  &=& \frac{Q} {r_{i}} ~.\label{eq8}
\end{eqnarray}

We should be noted that in the extended phase space the ADM mass can be treated as  the total 
gravitational  enthalpy of the system i.e. $M=H=U+PV$. Where $U$ is the thermal energy of the 
system\cite{kastor09}. Then the first law of BH thermodynamics in the extended phase space becomes
\begin{eqnarray}
dH &=&  T_{i} d{\cal S}_{i} + V_{i} dP +\Phi_{i} dQ ~. \label{eq9}
\end{eqnarray}
and the corresponding Smarr-Gibbs-Duhem relation becomes
\begin{eqnarray}
H &=&  2T_{i} {\cal S}_{i} - 2P V_{i} +Q \Phi_{i} ~. \label{eq10}
\end{eqnarray}
Now we compute the mass(or enthalpy) independent entropy sum and entropy product formula 
using Eqs.(\ref{eq1},\ref{eq2},\ref{eq4},\ref{eq5},\ref{eq6}):
\begin{eqnarray}
\sum_{i=1}^{4} \sqrt{{\cal S}_{i}} &=& 0 ~.\label{eq11}\\
\sum_{1\leq i<j\leq 4} \sqrt{{\cal S}_{i}{\cal S}_{j}} &=& \frac{3}{8P} ~.\label{eq12}\\
\prod_{i=1}^{4} \sqrt{{\cal S}_{i}} &=& \frac{3 \pi Q^2}{8P} ~.\label{eq13}
\end{eqnarray}
In terms of two horizons, the mass-independent entropy product formula should read
\begin{eqnarray}
 \frac{\left(\frac{3Q^2}{8P} \right)}{\sqrt{{\cal S}_{1}{\cal S}_{2}}}-
 \frac{{\cal S}_{1}+{\cal S}_{2}+\sqrt{{\cal S}_{1}{\cal S}_{2}}}{\pi} &=& \frac{3}{8\pi P} 
~. \label{eqq1}
\end{eqnarray}
Although it is a complicated function of two horizons but it is explicitly mass independent 
function of inner horizon and outer horizons.

Now we turn into another important relations that is the \emph{volume sum} and 
\emph{volume product} which are mass independent:
\begin{eqnarray}
\sum_{i=1}^{4} {V_{i}}^{\frac{1}{3}} &=& 0 ~.\label{eq14}\\
\sum_{1\leq i<j\leq 4}(V_{i}V_{j})^{\frac{1}{3}}  &=& \left(\frac{3}{32 \pi}\right)^{\frac{1}{3}}
\frac{1}{P} ~.\label{eq15}\\
\prod_{i=1}^{4} (V_{i})^{\frac{1}{3}}&=&  
\left(\frac{\pi}{6}\right)^{\frac{1}{3}}\frac{Q^2}{P} ~.\label{eq16}
\end{eqnarray}
Again in terms of two horizons, the mass independent volume product formula should 
be 
\begin{eqnarray}
\left(\frac{3}{32 \pi}\right)^{\frac{1}{3}} \frac{\left(\frac{Q^2}{P}\right)}{(V_{1}V_{2})^{\frac{1}{3}}}
-\left(\frac{3}{4\pi}\right)^{\frac{2}{3}}\left[{V_{1}}^{\frac{2}{3}}+{V_{2}}^{\frac{2}{3}}+(V_{1}V_{2})^{\frac{1}{3}}
\right]  &=& \frac{3}{8\pi P} 
~. \label{eqq2}
\end{eqnarray}
These are explicitly mass-independent relations in the extended phase space. 

Finally, the equation of state in the extended phase space should read 
\begin{eqnarray}
P &=& \frac{T_{i}} {2r_{i}} -\frac{1}{8\pi r_{i}^2}+\frac{Q^2}{8\pi r_{i}^4} ~. \label{eq17}
\end{eqnarray}
where $r_{i}=\left(\frac{3V_{i}}{4\pi}\right)^{1/3}$. Now in terms of specific volume 
$v_{i}=2r_{i}$ the above equation could be re-written as 
\begin{eqnarray}
P &=& \frac{T_{i}} {v_{i}} -\frac{1}{2\pi v_{i}^2}+\frac{2Q^2}{\pi v_{i}^4} ~. \label{eq18}
\end{eqnarray}
The critical point can be obtained from the following conditions:
\begin{eqnarray}
\frac{\partial P}{\partial v_{i}} &=& \frac{\partial^2 P}{\partial v_{i}^2}=0  ~. \label{eq19}
\end{eqnarray}
The critical values are explicitly computed in \cite{david12}.
Defining further $p=\frac{P}{P_{c}}$, $\nu_{i}=\frac{v_{i}}{v_{c}}$ and $\tau_{i}=\frac{T_{i}}{T_{c}}$, 
the law of corresponding states become 
\begin{eqnarray}
8 \tau_{i} &=& 3\nu_{i}\left(p+\frac{2}{\nu_{i}^2}\right)-\frac{1}{\nu_{i}^3}  ~. \label{eq20}
\end{eqnarray}
\footnote{The critical values for charged-AdS BH determined in \cite{david12} are $P_{c}=\frac{1}{96\pi Q^2}$, 
$v_{c}=2\sqrt{6} Q$ and $T_{c}=\frac{\sqrt{6}}{18\pi Q}$}.
In the extended phase space the Gibbs free energy could be defined as 
\begin{eqnarray}
 G_{i} &=& H-T_{i}S_{i}=M-T_{i}S_{i}=\frac{r_{i}}{4}-\frac{2\pi P}{3} r_{i}^3+\frac{3Q^2}{4r_{i}} ~. \label{eq21}
\end{eqnarray}
and the Helmholtz free energy is given by 
\begin{eqnarray}
 F_{i} &=& G_{i}-PV_{i}=\frac{r_{i}}{2}-\pi T_{i} r_{i}^2+\frac{Q^2}{2r_{i}} ~. \label{eq22}
\end{eqnarray}
which is very important to determine the behavior of the critical exponents. 
\footnote{The critical exponents for the BH system are $\alpha$, $\beta$, $\gamma$ and $\delta$. The numerical values  
for charged-AdS BH are calculated in\cite{david12} as $\alpha=0$, $\beta=1/2$, $\gamma=1$ and $\delta=3$. }
One may compute the entropy via the relation:
\begin{eqnarray}
{\cal S}_{i} &=& - \left(\frac{\partial F_{i}}{\partial T_{i}}\right)_{V_{i}}=\pi r_{i}^2   ~. \label{eq23}
\end{eqnarray}
which is exactly same as in Eq. (\ref{eq5}). There are two types of specific heat. The specific heat 
at constant thermodynamic volume and the specific heat at constant pressure. They are defined as 
\begin{eqnarray}
\left( C_{V} \right)_{i} &=& T_{i} \left(\frac{\partial S_{i}}{\partial T_{i}}\right)_{V}   .~\label{cv}
\end{eqnarray}
and
\begin{eqnarray}
\left( C_{P} \right)_{i} &=& T_{i} \left(\frac{\partial S_{i}}{\partial T_{i}}\right)_{P}
.~\label{cp}
\end{eqnarray}
From Eq. \ref{eq23}, we can easily see that the entropy ${\cal S}_{i}$ is independent of $T_{i}$ therefore we get 
\begin{eqnarray}
\left( C_{V} \right)_{i} &=& 0  .~\label{cv1}
\end{eqnarray}
and we find 
\begin{eqnarray}
\left( C_{P} \right)_{i} &=& 
-2\pi r_{i}^2 \frac{\left(1-\frac{Q^2}{r_{i}^2}+8\pi Pr_{i}^2\right)}{\left(1-\frac{3Q^2}{r_{i}^2}-8\pi Pr_{i}^2\right)} 
.~\label{cp2}
\end{eqnarray}
The specific heat $C_{P}$ diverges at  
\begin{eqnarray}
8\pi P r_{i}^4-r_{i}^2+3Q^2 &=& 0 ~. \label{eq26}
\end{eqnarray}
or i.e. at
\begin{eqnarray}
r_{i} &=& \pm \sqrt{\frac{1\pm\sqrt{1-96\pi Q^2 P}}{16\pi P}} ~. \label{eq27}
\end{eqnarray}
which implies a second order phase transition occurs at that point.


\section{\label{quint} Thermodynamic properties of the RN-AdS BH surrounded by quintessence:}

In this section, we will show how the quintessence dark energy matter does affect on the 
thermodynamic product relation in the extended phase space. The metric function of Eq. (\ref{metric}) 
for  RN-AdS BH surrounded by quintessence can be written as \cite{kies03}
\begin{eqnarray}
{\cal U}(r) &=&  1-\frac{2M}{r}+\frac{Q^2}{r^2}-\frac{a}{r^{3w_{q}+1}}-\frac{\Lambda}{3}r^2
~. \label{eq28}
\end{eqnarray}
where $w_{q}$ is the state parameter and $a$ is the normalization factor related to the density 
of quintessence. The ranges for quintessence dark energy is $-1<w_{q}<-\frac{1}{3}$ and for 
phantom dark energy: $w_{q}<1$. In terms of $a$, the density of quintessence can be defined as
\begin{eqnarray}
\rho_{q} &=& -\frac{3aw_{q}}{2r^{3w_{q}+1}} ~. \label{eq29}
\end{eqnarray}

In the extended phase space the function can be written as 
\begin{eqnarray}
{\cal U}(r) &=&  1-\frac{2M}{r}+\frac{Q^2}{r^2}-\frac{a}{r^{3w_{q}+1}}+\frac{8\pi P}{3}r^2
~. \label{eq30}
\end{eqnarray}

Now the horizon Eq. can be written as  
\begin{eqnarray}
\frac{8\pi P}{3}r^{3w_{q}+3}+r^{3w_{q}+1}-2Mr^{3w_{q}}+Q^2r^{3w_{q}-1} -a &=& 0 ~.\label{eq31}
\end{eqnarray}
Using  Vieta's theorem, we find 
\begin{eqnarray}
\sum_{i=1}^{3w_{q}+3} r_{i} &=& 0 ~.\label{eq32}\\
\sum_{1\leq i<j\leq (3w_{q}+3)} r_{i}r_{j} &=& \frac{3}{8\pi P} ~.\label{eq33}\\
\sum_{1\leq i<j<k\leq (3w_{q}+3)} r_{i}r_{j} r_{k} &=&  \frac{3M}{4\pi P} ~.\label{eq34}\\
\sum_{1\leq i<j<k<l\leq (3w_{q}+3) } r_{i}r_{j} r_{k}r_{l} &=&  \frac{3Q^2}{8\pi P} ~.\label{eq35}\\
\prod_{i=1}^{3w_{q}+3} r_{i} &=&  a ~.\label{eq36}
\end{eqnarray}
It should be mentioned that $3w_{q}$ is an integer quantity.

The entropy ${\cal S}_{i}$ and electric potential $\Phi_{i}$ are same as in RN-AdS case. Now the mass of 
the BH could be expressed in terms of the horizon radius and dynamic pressure:
\begin{eqnarray}
M &=& \frac{r_{i}}{2}\left(1+\frac{Q^2}{r_{i}^2}-\frac{a}{r_{i}^{3w_{q}+1}}+\frac{8\pi P}{3}r_{i}^2\right)
~. \label{eq37}
\end{eqnarray}
Hence the first law of thermodynamics becomes
\begin{eqnarray}
dH &=&  T_{i} d{\cal S}_{i} + V_{i} dP +\Phi_{i} dQ + {\cal A}_{i} da~. \label{eq38}
\end{eqnarray}
where ${\cal A}_{i}=\left(\frac{\partial H}{\partial a} \right)_{S_{i},Q,P}=-\frac{1}{2r_{i}^{3w_{q}}}$ is 
defined to be a physical quantity conjugate to the state parameter \cite{li}. 
The corresponding Smarr relation reads 
\begin{eqnarray}
H &=&  2T_{i} {\cal S}_{i} - 2P V_{i} +Q \Phi_{i}+(1+3w_{q}){\cal A}_{i} da ~. \label{eq39}
\end{eqnarray}
Again the mass(or enthalpy) independent entropy sum and entropy product relations are
\begin{eqnarray}
\sum_{i=1}^{(3w_{q}+3)} \sqrt{{\cal S}_{i}} &=& 0 ~.\label{eq40}\\
\sum_{1\leq i<j\leq (3w_{q}+3)} \sqrt{{\cal S}_{i}{\cal S}_{j}} &=& \frac{3}{8P} ~.\label{eq41}\\
\prod_{i=1}^{(3w_{q}+3)} \sqrt{\frac{{\cal S}_{i}}{\pi}} &=& a ~.\label{eq42}
\end{eqnarray}
Similarly, the mass independent volume sum and volume product relations are 
\begin{eqnarray}
\sum_{i=1}^{(3w_{q}+3)} {V_{i}}^{\frac{1}{3}} &=& 0 ~.\label{eq43}\\
\sum_{1\leq i<j\leq (3w_{q}+3)}(V_{i}V_{j})^{\frac{1}{3}}  &=& \left(4\pi\right)^{\frac{2}{3}} \frac{3}{8\pi P}
~.\label{eq44}\\
\prod_{i=1}^{(3w_{q}+3)} \left(\frac{3V_{i}}{4\pi}\right)^{\frac{1}{3}}&=&  a ~.\label{eq45}
\end{eqnarray}
These are explicitly mass-independent relations for RN-AdS BH surrounded by quintessence.
It follows from the above analysis that the entropy product and volume product relations are 
strictly dependent on \emph{quintessence dark energy matter}. It is quite interesting to mentioned 
that the entropy product is mass-independent but there has been effect of quintessence dark energy matter on
that thermodynamic product relations. 

Now the BH temperature for all the horizons could be defined as 
\begin{eqnarray}
T_{i} &=& \frac{1}{4\pi r_{i}} \left(1+8\pi P r_{i}^2-\frac{Q^2}{r_{i}^2}+ \frac{3aw_{q}}{r_{i}^{3w_{q}+1}}\right)
~. \label{eq46}
\end{eqnarray} 
and the BH equation of state should read
\begin{eqnarray}
P &=& \frac{T_{i}} {2r_{i}} -\frac{1}{8\pi r_{i}^2}+\frac{Q^2}{8\pi r_{i}^4}-
\frac{3aw_{q}}{8\pi r_{i}^{3(w_{q}+1)}} ~. \label{eq47}
\end{eqnarray}
where $r_{i}=\left(\frac{3V_{i}}{4\pi}\right)^{1/3}$. Again in terms of specific 
volume $v_{i}=2r_{i}$ the above equation should be rewritten as 
\begin{eqnarray}
P &=& \frac{T_{i}} {v_{i}} -\frac{1}{2\pi v_{i}^2}+\frac{2Q^2}{\pi v_{i}^4}-\frac{3aw_{q}2^{4w_{q}}}{\pi v_{i}^{3(w_{q}+1)}}
~. \label{eq48}
\end{eqnarray}
The critical values are explicitly calculated in \cite{li}. So we do not written here. 
The Gibbs free energy for all the horizons could be written as 
\begin{eqnarray}
 G_{i} &=& H-T_{i}S_{i}=M-T_{i}S_{i}
 =\frac{r_{i}}{4}-\frac{2\pi P}{3} r_{i}^3+\frac{3Q^2}{4r_{i}}-\frac{3aw_{q}+2a}{4 r_{i}^{3w_{q}}} 
 ~. \label{eq49}
\end{eqnarray}
Again we compute the specific heat at constant thermodynamic pressure to study the local stability of the BH given by
\begin{eqnarray}
\left(C_{P} \right)_{i} &=& -2\pi r_{i}^2 \frac{\left(1-\frac{Q^2}{r_{i}^2}+\frac{3aw_{q}}{r_{i}^{3w_{q}+1}}+8\pi Pr_{i}^2\right)}
{\left(1-\frac{3Q^2}{r_{i}^2}+\frac{3(2+3w_{q})aw_{q}}{r_{i}^{3w_{q}+1}}-8\pi Pr_{i}^2\right)}  .~\label{sh1}
\end{eqnarray}
It should be noted that the specific heat diverges  at 
\begin{eqnarray}
1-\frac{3Q^2}{r_{i}^2}+\frac{3(2+3w_{q})aw_{q}}{r_{i}^{3w_{q}+1}}-8\pi Pr_{i}^2 &=& 0 .~\label{sh2}
\end{eqnarray}
which signals a second order phase transition.

\section{\label{fr} Thermodynamic properties of AdS BH in $f(R)$ gravity:}
This section is dedicated to study the thermodynamic properties of a static, spherically symmetric AdS BH in  $f(R)$ gravity. 
It is a kind of modified gravity and it is  very useful tool for explaining the current and future state of the 
accelerating universe. It is also helpful for explaining the inflation and structure formation in the early universe.
The metric \cite{moon11,chen} function for this kind of gravity can be written as 
\begin{eqnarray}
{\cal U}(r) &=&  1-\frac{2m}{r}+\frac{q^2}{\alpha r^2}-\frac{R_{0}}{12}r^2 ~. \label{eq50}
\end{eqnarray}
where $\alpha=1+f'(R_{0})$. The quantities $m$ and $q$ are related to the $M$(ADM mass) and $Q$(electric charge) in 
this gravity becomes
\begin{eqnarray}
M =m\alpha ,\,\,\,\, Q=\frac{q}{\sqrt{\alpha}}~. \label{eq51}
\end{eqnarray}
As is the thermodynamic pressure in $f(R)$ gravity can be written as $P=-\frac{\Lambda}{8\pi} \alpha$ and the 
constant scalar curvature  as $R_{0}=-\frac{12}{\ell^2}=4\Lambda$. The corresponding thermodynamic volume can be 
defined as 
$V_{i} = \frac{4}{3}\pi r_{i}^3$.
Therefore the horizon equation should read 
\begin{eqnarray}
\frac{8\pi P}{3}\alpha r^4+\alpha r^2-2m\alpha r+q^2 &=& 0 ~.\label{eq52}
\end{eqnarray}
To finding the roots we again apply the Vieta's rule, we have 
\begin{eqnarray}
\sum_{i=1}^{4} r_{i} &=& 0 ~.\label{eq53}\\
\sum_{1\leq i<j\leq 4} r_{i}r_{j} &=& \frac{3}{8\pi P} ~.\label{eq54}\\
\sum_{1\leq i<j<k\leq 4} r_{i}r_{j} r_{k} &=&  \frac{3m}{4\pi P} ~.\label{eq55}\\
\prod_{i=1}^{4} r_{i} &=& \frac{3q^2}{8\pi \alpha P} ~.\label{eq56}
\end{eqnarray}

The entropy can be defined as 
\begin{eqnarray}
{\cal S}_{i}  &=& \pi \alpha  r_{i}^2 ~.\label{eq57}
\end{eqnarray}
and the BH temperature\cite{chen} should be 
\begin{eqnarray}
T_{i} &=&  \frac{1}{4\pi r_{i}} \left(1+\frac{8\pi P}{\alpha} r_{i}^2-\frac{q^2}{\alpha r_{i}^2} \right)
~.\label{eq58}
\end{eqnarray} 
Again the electric potential in $f(R)$ gravity could be defined as 
\begin{eqnarray}
\Phi_{i}  &=& \frac{q}{r_{i}}\sqrt{\alpha} ~.\label{eq59}
\end{eqnarray}

Now the mass(or enthalpy) independent entropy sum and entropy product formula in $f(R)$ 
gravity should read:
\begin{eqnarray}
\sum_{i=1}^{4} \sqrt{{\cal S}_{i}} &=& 0 ~.\label{eq60}\\
\sum_{1\leq i<j\leq 4} \sqrt{{\cal S}_{i}{\cal S}_{j}} &=& \frac{3\alpha}{8P} ~.\label{eq61}\\
\prod_{i=1}^{4} \sqrt{{\cal S}_{i}} &=& \frac{3 \pi\alpha q^2}{8P} ~.\label{eq62}
\end{eqnarray}
In terms of two horizons, the mass-independent entropy product formula reads as
\begin{eqnarray}
 \frac{\left(\frac{3q^2}{8P} \right)}{\sqrt{{\cal S}_{1}{\cal S}_{2}}}-
 \frac{{\cal S}_{1}+{\cal S}_{2}+\sqrt{{\cal S}_{1}{\cal S}_{2}}}{\pi \alpha} &=& \frac{3}{8\pi P} 
~. \label{eqq3}
\end{eqnarray}
Again the  mass independent \emph{volume sum} and \emph{volume product} becomes
\begin{eqnarray}
\sum_{i=1}^{4} {V_{i}}^{\frac{1}{3}} &=& 0 ~.\label{eq63}\\
\sum_{1\leq i<j\leq 4}(V_{i}V_{j})^{\frac{1}{3}}  &=& \left(\frac{3}{32 \pi}\right)^{\frac{1}{3}}\frac{1}{P} 
~.\label{eq64}\\
\prod_{i=1}^{4} (V_{i})^{\frac{1}{3}}&=&  
\left(\frac{\pi}{6}\right)^{\frac{1}{3}} \frac{q^2}{\alpha P} ~.\label{eq65}
\end{eqnarray}
Again in terms of two horizons, the mass independent volume product formula reads
\begin{eqnarray}
\left(\frac{3}{32 \pi}\right)^{\frac{1}{3}} \frac{\left(\frac{q^2}{P}\right)}{\alpha(V_{1}V_{2})^{\frac{1}{3}}}
-\left(\frac{3}{4\pi}\right)^{\frac{2}{3}}\left[{V_{1}}^{\frac{2}{3}}+{V_{2}}^{\frac{2}{3}}+(V_{1}V_{2})^{\frac{1}{3}}
\right]  &=& \frac{3}{8\pi P} ~. \label{eqq4}
\end{eqnarray}
Once again these are explicitly mass-independent relation in the extended phase space in $f(R)$ gravity. It should be noted 
that in the limit $\alpha=1$, one obtains the result of RN-AdS BH in extended phase space. For our record we should 
be noted that the equation of state in $f(R)$ gravity\cite{chen}: 
\begin{eqnarray}
P &=& \frac{\alpha T_{i}} {2r_{i}} -\frac{\alpha}{8\pi r_{i}^2}+\frac{q^2}{8\pi r_{i}^4} ~. \label{eq66}
\end{eqnarray}
In terms of specific volume $v_{i}=2r_{i}$, the above Eq. could be rewritten as 
\begin{eqnarray}
P &=& \frac{\alpha T_{i}} {v_{i}} -\frac{\alpha}{2\pi v_{i}^2}+\frac{2q^2}{\pi v_{i}^4} ~. \label{eq67}
\end{eqnarray}
From the equation of state we can easily derived the critical constants by applying the appropriate condition. 
\footnote{The critical values for $f(R)$ gravity explicitly computed in \cite{chen} are $P_{c}=\frac{\alpha^2}{96\pi q^2}$, 
$v_{c}=\frac{2q\sqrt{6}}{\sqrt{\alpha}}$ and $T_{c}=\frac{\sqrt{6 \alpha}}{18\pi q}$}.

Finally, the Gibbs free energy \cite{chen} should read
\begin{eqnarray}
 G_{i} &=& \frac{\alpha r_{i}}{4}-\frac{2\pi P}{3} r_{i}^3+\frac{3q^2}{4r_{i}} ~. \label{eq68}
\end{eqnarray}

In this case, the specific heat at constant pressure is found to be
\begin{eqnarray}
\left(C_{P}\right)_{i} &=& -2\pi r_{i}^2 \frac{\left(1-\frac{q^2}{\alpha r_{i}^2}+8\pi Pr_{i}^2\right)}
{\left(1-\frac{3q^2}{\alpha r_{i}^2}-8\pi Pr_{i}^2\right)}  .~\label{sh3}
\end{eqnarray}
The specific heat diverges  at 
\begin{eqnarray}
8\pi \alpha P r_{i}^4-\alpha r_{i}^2+3q^2 &=& 0 ~. \label{eq69}
\end{eqnarray}
or
i.e. at
\begin{eqnarray}
r_{i} &=& \pm \sqrt{\frac{\alpha \pm\sqrt{\alpha^2-96\pi \alpha Pq^2}}{16\pi \alpha P}} ~. \label{eq70}
\end{eqnarray}
Again it signals a second order phase transition. 


\section{\label{dis} Conclusion:}
The present study demonstrated that the thermodynamic properties of spherically symmetric charged-AdS black 
hole, charged AdS BH surrounded by quintessence and charged AdS BH in $f(R)$ gravity in the extended 
phase-space. The extended phase space means where the cosmological constant should be treated as 
thermodynamic pressure and its conjugate variable as a thermodynamic volume. We  derived various 
thermodynamic products particularly entropy products and thernodynamic volume products. In all the three cases, 
it has been shown that the mass(or enthalpy) independent properties turn out to be an universal like quantities. It should 
be noted that the presence  of the quintessence matter does affect on the expression of entropy product 
and thermodynamic volume products. The first law of BH thermodynamics and Smarr formula have been studied
for all the horizons. The BH equation of state has been derived for all the horizons. The divergence of the 
specific heat indicates that the second order phase transition should occur at a certain condition. In summary, the 
thermodynamic relations that we derived provide some universal characterization of the BH which \emph{may} provide 
insight into the origin of BH entropy both \emph{inner and outer}. 

\appendix

\section{}
In the main work, we have considered the thermodynamic volume for different spherically symmetric cases where it is 
related to the entropy via a proportionality of the type $V_{i} \propto r_{i} {\cal S}_{i}$ and we proved that for 
each cases, the thermodynamic product relations are \emph{independent of mass} thus the relations are  \emph{universal} 
in nature. Now here we shall show what happens in case of axisymmetric cases where the thermodynamic volume is 
\emph{not} directly proportional to the above mentioned relation? For example, we have taken the Kerr BH.  
The thermodynamic quantities for all the horizons ${\cal H}^{\pm}$ \cite{ppepjc} are 
\begin{eqnarray}
r_{\pm} = M \pm \sqrt{M^{2}-a^2}, \,\, {\cal A}_{\pm} = 4 \pi \left(r_{\pm}^2+a^2\right), \,\, \\
S_{\pm} = \pi \left(r_{\pm}^2+a^2 \right),\,\, T_{\pm} =\frac{r_{\pm}-r_{\mp}}{4\pi (r_{\pm}^2+a^2)}, \\
\Omega_{\pm} = \frac{a}{2Mr_{\pm}} ~.\label{ap}
\end{eqnarray}
and the thermodynamic volume \cite{nata} for ${\cal H}^{+}$ is
\begin{eqnarray}
V_{+} &=& \frac{{\cal A}_{+} r_{+}}{3} \left[1+\frac{a^2}{2r_{+}^2} \right] ~.\label{ap1}
\end{eqnarray} 
and we claim that the thermodynamic volume  for ${\cal H}^{-}$ is defined to be
\begin{eqnarray}
V_{-} &=& \frac{{\cal A}_{-} r_{-}}{3} \left[1+\frac{a^2}{2r_{-}^2} \right] ~.\label{vi}
\end{eqnarray} 
We find an interesting relation between volume products and area products  for an axisymmetric 
spacetime having two physical horizons namely ${\cal H}^{+}$ and ${\cal H}^{-}$:
\begin{eqnarray}
V_{+} V_{-} &=& \frac{{\cal A}_{+}{\cal A}_{-}}{18} \left(r_{+}+r_{-} \right)^2 ~.\label{vi1}
\end{eqnarray}
The main potential point of interest here is that the product of thermodynamic volume of ${\cal H}^{\pm}$ 
and it is found to be for Kerr BH:
\begin{eqnarray}
V_{+} V_{-} &=& \frac{128}{9} \pi^2 J^2 M^2  ~.\label{ap2}
\end{eqnarray}
It indicates that the thermodynamic volume product  does \emph{depend} on the mass parameter and the universality that 
we have found in spherically symmetric cases  breaks down for axisymmetric cases. Thus the conclusion is that although 
the area (or entropy product ) of ${\cal H}^{\pm}$ is universal for simple Kerr BH but the volume product is 
\emph{not universal}. The result should be valid for KN BH as well and the volume product is found to be 
\begin{eqnarray}
V_{+} V_{-} &=& \frac{128}{9} \pi^2 \left(J^2+\frac{Q^4}{4}\right) M^2  ~.\label{ap3}
\end{eqnarray}
Therefore, it can be easily extend to Kerr-AdS BH and KN-AdS BH also.


\section{}
In this section, we shortly introduce $P-V$ criticality of Cauchy horizon  for charged AdS BH. Does the inner 
horizon obey an equation of state? What happens to the inner horizons during phase transition has been 
discussed in \cite{1507}. Here we are interested to show what happens the BH equation of state in case of Cauchy 
horizon? What are the values of critical constant for this horizon does it same as is for event horizon. This is the 
main interest here. Since we have taken charged AdS space-time and we have assumed that the BH should have at least 
two physical horizons. The outer horizon $(r_{+})$ and inner horizon $(r_{-})$. Then we find the relevant 
thermodynamic quantities for ${\cal H}^{+}$ \cite{david12}:
$$
{\cal A}_{+}  = 4\pi r_{+}^2, \,\, {\cal S}_{+}  = \pi r_{+}^2, \,\, \Phi_{+} = \frac{Q} {r_{+}},
V_{+} = \frac{4}{3}\pi r_{+}^3,\,\,  T_{+} =\frac{1}{4\pi r_{+}} \left(1+8\pi P r_{+}^2-\frac{Q^2}{r_{+}^2} \right)
$$
$$
G_{+} =M-T_{+}S_{+}=M-\pi r_{+}^2 T_{+},\,\, F_{+} = G_{+}-PV_{+}, \,\, dH = T_{+} d{\cal S}_{+} + V_{+} dP+\Phi_{+} dQ,\,\,
$$
$$
P = \frac{T_{+}} {2r_{+}} -\frac{1}{8\pi r_{+}^2}+\frac{Q^2}{8\pi r_{+}^4}, \,\, r_{+}=\left(\frac{3V_{+}}{4\pi}\right)^{1/3}
$$
\begin{eqnarray}
P = \frac{T_{+}} {v_{+}} -\frac{1}{2\pi v_{+}^2}+\frac{2Q^2}{\pi v_{+}^4}, \,\, v_{+}=2r_{+}
~.\label{ap4}
\end{eqnarray}
The critical point can be obtained at the point of inflection of ${\cal H}^{+}$:
\begin{eqnarray}
\frac{\partial P}{\partial v_{+}} &=& \frac{\partial^2 P}{\partial v_{+}^2}=0  ~. \label{ap5}
\end{eqnarray}
and the critical values explicitly derived in \cite{david12} are 
\begin{eqnarray}
P_{c}^{+} &=& \frac{1}{96\pi Q^2}, \,\, v_{c}^{+}=2\sqrt{6} Q,\,\, T_{c}^{+}=\frac{\sqrt{6}}{18\pi Q}~. \label{ad5}
\end{eqnarray}

Now we have derived the thermodynamic quantities for ${\cal H}^{-}$:
$$
{\cal A}_{-}  = 4\pi r_{-}^2, \,\, {\cal S}_{-}  = \pi r_{-}^2, \,\, \Phi_{-} = \frac{Q} {r_{-}},
V_{-} = \frac{4}{3}\pi r_{-}^3,\,\,  T_{-} =-\frac{1}{4\pi r_{-}} \left(1+8\pi P r_{-}^2-\frac{Q^2}{r_{-}^2} \right)
$$
$$
G_{-} =-M-T_{-}S_{-}=-M-\pi r_{-}^2 T_{-},\,\, F_{-} = G_{-}-PV_{-}, \,\, -dH = -T_{-} d{\cal S}_{-} + V_{-} dP+\Phi_{-} dQ,\,\,
$$
$$
P = -\frac{T_{-}} {2r_{-}} -\frac{1}{8\pi r_{-}^2}+\frac{Q^2}{8\pi r_{-}^4}, \,\, r_{-}=\left(\frac{3V_{-}}{4\pi}\right)^{1/3}
$$
\begin{eqnarray}
P = -\frac{T_{-}} {v_{-}} -\frac{1}{2\pi v_{-}^2}+\frac{2Q^2}{\pi v_{-}^4}, \,\, v_{-}=2r_{-}
~.\label{ap6}
\end{eqnarray}
Similarly, we can compute the critical values for ${\cal H}^{-}$  by applying the condition 
at the point of inflection:
\begin{eqnarray}
\frac{\partial P}{\partial v_{-}} &=& \frac{\partial^2 P}{\partial v_{-}^2}=0  ~. \label{ap7}
\end{eqnarray}
and the critical values we find
\begin{eqnarray}
P_{c}^{-} &=& \frac{1}{96\pi Q^2}, \,\, v_{c}^{-} =2\sqrt{6} Q,\,\, \mbox{and} \,\, T_{c}^{-} =-\frac{\sqrt{6}}{18\pi Q}~. 
\label{ap8}
\end{eqnarray}
The only difference in critical values between two horizons is 
\begin{eqnarray}
T_{c} =-\frac{\sqrt{6}}{18\pi Q}~. \label{ap9}
\end{eqnarray}
The critical temperature for ${\cal H}^{-}$ is negative but other values are same. It is also true that 
the $P-V$ diagram is qualitatively different for ${\cal H}^{+}$ and ${\cal H}^{-}$. The critical ratio for 
${\cal H}^{-}$ is calculated to be 
\begin{eqnarray}
\rho_{c}^{-} &=& \frac{P_{c} v_{c}}{T_{c}}=-\frac{3}{8} ~. \label{cr}
\end{eqnarray}
where as for ${\cal H}^{+}$, $\rho_{c}^{+}=\frac{3}{8}$. Thus, we conclude that 
\begin{eqnarray}
P_{c}^{-} &=& P_{c}^{+}\\
v_{c}^{-} &=& v_{c}^{+} \\
T_{c}^{-} &=& -T_{c}^{+} ~. \label{tcpm}\\
\rho_{c}^{-} &=& -\rho_{c}^{+} ~. \label{tcpm1} 
\end{eqnarray}

Finally, using the properties of symmetry in nature of $r_{\pm}$, one obtains the following thermodynamic 
quantities at  ${\cal H}^{-}$:
\begin{eqnarray}
{\cal A}_{-} &=& {\cal A}_{+}|_{r_{+}\leftrightarrow r_{-}}, {\cal S}_{-}={\cal S}_{+}|_{r_{+}\leftrightarrow r_{-}}, 
\Omega_{-}=\Omega_{+}|_{r_{+}\leftrightarrow r_{-}}, \Phi_{-}=\Phi_{+}|_{r_{+}\leftrightarrow r_{-}} \\
T_{-} &=& -T_{+}|_{r_{+}\leftrightarrow r_{-}}, \, V_{-} = V_{+}|_{r_{+}\leftrightarrow r_{-}}, 
G_{-} = -G_{+}|_{r_{+}\leftrightarrow r_{-}} ~.\label{ap10}
\end{eqnarray}

\section*{Acknowledgements}

I  would like to thank the Editor Prof. J. P. S. Lemos and the anonymous Referee for their useful suggestions for which 
the manuscript is enhanced substantially. I am specially thanks to the Referee for some interesting comments.


\end{document}